\documentclass[%
    twocolumn,
    amsmath,
    amssymb,
    prl,
    aps,
    longbibliographystyle,
    nofootinbib,
    superscriptaddress,
    preprintnumbers,
]{revtex4-2}
\pdfoutput=1
\usepackage{cancel,multirow,graphicx, nicefrac}
\usepackage{dcolumn}
\usepackage[dvipsnames]{xcolor}
\usepackage{hyperref}
\hypersetup{%
    colorlinks=true,
    linkcolor=SeaGreen,
    urlcolor=Periwinkle,
    citecolor=Periwinkle
}
\usepackage[nameinlink]{cleveref}
\usepackage{nicefrac}

\usepackage{tikz}
\definecolor{lime}{HTML}{A6CE39}
\DeclareRobustCommand{\orcidicon}{\hspace{-1mm}
	\begin{tikzpicture}
	\draw[lime, fill=lime] (0,0) 
	circle [radius=0.16] 
	node[white] {{\fontfamily{qag}\selectfont \tiny \,ID}};
	\draw[white, fill=white] (-0.0525,0.095) 
	circle [radius=0.007];
	\end{tikzpicture}
	\hspace{-3mm}
}
\foreach \x in {A, ..., Z}{\expandafter\xdef\csname orcid\x\endcsname{\noexpand\href{https://orcid.org/\csname orcidauthor\x\endcsname}
			{\noexpand\orcidicon}}
}

\def\tkspn{\tau^- \to K_s\pi^- \nu}

\begin{document}
\preprint{UT-WI-08-2026}
\title{Constraining ALP-Meson overlaps from \texorpdfstring{\(K\pi\)}{K-pi} form factors}

\author{Triparno Bandyopadhyay\orcidA}
 \email{gondogolegogol@gmail.com}
 \affiliation{Department of Physics and Nanotechnology, SRM Institute of Science and Technology, Kattankulathur, Tamil Nadu 603203, India.}
\author{Subhajit Ghosh\orcidB}%
\email{sghosh@utexas.edu}
\affiliation{%
Texas Center for Cosmology and Astroparticle Physics, Weinberg Institute,\\
Department of Physics, The University of Texas at Austin, Austin, TX 78712, USA
}%
\date{\today}

\begin{abstract} 
    We present the first constraints on the overlaps between an Axion-like
    particle (ALP) and the $\pi^0$ and $\eta$ mesons from the analysis of the
    distortions to the $\langle K|\overline{s}\gamma^\mu u | \pi\rangle$ form
    factors.  We demonstrate that these distortions can be tightly constrained by
    combining data from $\tau^-\to \pi^0 K^-\nu$ and $K^+\to \pi^0\ell^+\nu$
    decays, and go on to map the constraints to the ALP-meson overlaps. 
    We establish that, in general, the ALP-meson and meson-ALP overlaps are
    different due to the presence of ALP-quark derivative couplings in the UV
    Lagrangian, and need to be treated separately. 
    %
    Using lattice results and BaBar, Belle, and NA48/2 data, we obtain
    exclusion limits on the overlaps and give projections for Belle II. Our
    techniques are independent of the branching ratios of the ALP, hence, robust
    against ALP decay channel assumptions.
    For masses of the ALP below 1~GeV, the bounds on the effective scale of the
    ALP physics extend to \(\mathcal{O}\)(10 TeV) for restricted regions of
    the parameter space for the ALP-$\pi$ and $\pi$-ALP overlaps. On the other
    hand, these bounds persist for extended regions of the parameter
    space for ALP-$\eta$ and $\eta$-ALP overlaps. 
\end{abstract}

\maketitle

\emph{Introduction:} Axion-like particles (ALPs), which emerge as the
pseudo-Nambu-Goldstone boson (pNGB) of a \(U(1)\) symmetry, spontaneously broken
at the UV, are among the most well-motivated new physics (NP)
scenarios~\cite{Peccei:1977hh, Peccei:1977ur, Weinberg:1977ma, Wilczek:1977pj,
tHooft:1976rip, Preskill:1982cy, Dine:1982ah, Abbott:1982af, Graham:2015cka,
Hook:2016mqo, Trifinopoulos:2022tfx, Co:2019wyp, Chakraborty:2021fkp}. Due to
their Goldstone-nature, ALPs can be naturally light, residing far below the
electroweak (EW) scale, and are at the forefront of NP searches below the GeV
scale.  Specifically, the extent to which the ALP (\(a\)) redefines the Standard
Model (SM) \(\pi^0\), \(\eta\), and their interactions, through \(a\)-\(\pi^0\)
and \(a\)-\(\eta\) overlaps in the flavor basis, are of great interest.  These overlaps are `rotated' away to go to the diagonal basis of the pions. The
`rotations' are general-linear transformations---due to the presence of both
kinetic and mass mixing terms---under which the path integral remains invariant.
These overlaps are sourced by the ALP couplings to the SM quarks, either
derivatively (\(\partial_\mu a \bar{f}\gamma^\mu f\)) or as phases in the Yukawa
interactions (\(\bar{f}_Le^{ia}f_R\))\footnote{After rotating away the
\(\protect{aG\tilde{G}}\) coupling, where \(G(G_{\mu\nu})\) is the gluon field
strength tensor.}, resulting in off-diagonal kinetic and mass terms,
respectively.

Existing bounds on \(a\)-\(\pi^0\) overlaps from direct production of \(a\) in the final (or initial) states are over a limited kinematic range due to the
experimental sensitivity to ALP decay modes.~(see, \emph{e.g.},
\cite{Altmannshofer:2019yji}). They
are critically dependent on the branching ratios~(BR) and the mass of the ALP,
while bounds are completely missing for \(a\)-\(\eta\) overlaps due to the
scarcity of data involving the \(\eta\) meson\footnote{Check
Ref.~\cite{Gao:2022xqz} for a recent analysis on the lattice for a particular
model.}. To largely circumvent these issues, in this work, we present an
\emph{indirect} method of probing the ALP-mass--overlap plane by analyzing
\(K\pi\) form factors (FFs) over the full kinematic range corresponding to
\(\tau^-\to K\pi\nu\) and the truncated range corresponding to
\(K^\pm\to\pi^0\ell^\pm\nu\) (\(K_{\ell_3}, l = e,\mu\)) decays. To be precise, we analyze
precision measurements of the \(K^+\to\pi^0\ell^+\nu_\ell\) decay distribution
by the NA48/2 collaboration~\cite{NA482:2018rgv,
madigozhin_dmitry_2019_3560600}, \(\tau^-\to K^-\pi^0\nu_\tau\) decay
distributions by the BaBar collaboration~\cite{BaBar:2007yir}, \(\tau^-\to
K^0\pi^-\nu_\tau\) by the Belle collaboration~\cite{Belle:2007goc} with inputs
from the lattice computation of the \(K\pi\) FF by the European Twisted Mass
(ETM) collaboration~\cite{Carrasco:2016kpy}, and partial widths and BR
measurements of the decays~\cite{ParticleDataGroup:2024cfk}.

In the following, we describe the modification of the meson chiral Lagrangian in
the presence of representative ALP-quark couplings~\cite{Georgi:1986df,
Bauer:2020jbp, Bauer:2021wjo, Bauer:2021mvw, Bandyopadhyay:2021wbb}, the
corresponding modifications to the FFs~\cite{Bandyopadhyay:2021wbb}, and the
limits on the overlaps obtained from our analysis of these modifications, before
concluding. 

\emph{The modified Chiral Lagrangian:} The most general Lagrangian of terms
quadratic in the ALP and meson (\(\pi^0, \eta\)) fields, in the chiral
Lagrangian, is: 
\begin{align}
    \mathcal{L} &= \partial_\mu{\Pi}^\dagger \frac{1}{2} {K}\,                
                    \partial^\mu{\Pi}
                - {\Pi}^\dagger \frac{B^2}{2} \mu {\Pi}\,
\end{align}
where \({K}\) and \({\mu}\) are dimensionless non-diagonal kinetic and mass
matrices respectively, \(B\) is the scale of the condensate of the order of the
\(\rho\) mass, below which we obtain the chiral Lagrangian, and \({\Pi} =
\begin{pmatrix} \hat{a}, \hat{\pi}^0, \hat{\eta} \end{pmatrix}^T\)\footnote{For
this work, we tacitly assume that the flavor basis is written after the
\(\protect{\eta^\prime}\) has been diagonalised.}. The \(\hat{~}\) indicates the
flavor basis.

The presence of both kinetic and mass mixing makes the transformation to the
canonically diagonal basis a general linear transformation, which we can
factorize into an upper triangular and an orthogonal transformation (QR
decomposition), the former diagonalizing the kinetic part and the latter
diagonalizing the resulting mass part. The net transformation is:
\begin{align}
    \label{eq:mix}
    \begin{pmatrix}
        \hat{a}\\\hat{\pi}^0\\\hat{\eta}
    \end{pmatrix}\!\! =\!\!
    \begin{pmatrix}
        1+C_{aa} \xi^2 & C_{a\pi}\, \xi &C_{a\eta}\, \xi\\
        C_{\pi a}\, \xi & 1+C_{\pi\pi} \xi^2 &C_{\pi\eta}\, \xi\\
        C_{\eta a}\, \xi & C_{\eta\pi}\, \xi & 1+ C_{\eta\eta} \xi^2
    \end{pmatrix}\!\!\!
    \begin{pmatrix}
        {a}\\{\pi}^0\\{\eta}
    \end{pmatrix},
\end{align}
where the Lagrangian is diagonal in the un-hatted basis, and \(\xi=f_\pi/f_a\),
with \(f_\pi\) the pion decay constant, and \(f_a=4\pi \Lambda\) is the ALP
decay constant, associated the scale \(\Lambda\) which is the scale of where the
UV physics giving rise to the ALP is integrated out. We expand till the second
power in \(\xi\). In this work, our goal is to find constraints on the
off-diagonal elements \(C_{a\pi}\), \(C_{a\eta}\), \(C_{\pi a}\), and \(C_{\eta
a}\). To do so, we first find out how these overlaps are associated with the
modifications to the charged-current FF \(\langle K^+|\overline{s}\gamma^\mu
u|\pi^0\rangle\).

In order to link the meson-ALP overlaps specifically to the charged current
decays of the \(K^\pm\) and the \(\tau^\pm\), we write down a simple UV scenario
with specific operators that affect these decays. This is done so as not
to obfuscate our results by an untractable number of Wilson Coefficients, as our main goal is to put forward the technique. We
concentrate on an ALP-quark Lagrangian that contains only \(t_8\),
isospin-breaking interactions, where \(t_8\) is the standard Gell-Mann matrix.
The ALP-quark Lagrangian at the EW symmetry-breaking scale
is taken to be~\cite{Bandyopadhyay:2021wbb}: 
\begin{align}
\label{eq:LagQ}
    &\mathcal{L} \supset C^{8}_{L}\frac{\partial^\mu a}{f_a} 
        \overline{q}_L \gamma^\mu t_{8}   q_L
       + iC_{LR}^8 \frac{a}{f_a}\overline{q}_L M t^8 q_R+\mathrm{h.c.}\;.
\end{align}
Here, \(C_i^j\) are the relevant Wilson coefficients, and
\(M=\mathrm{diag}(m_u,m_d,m_s)\). As we do not consider any current proportional
to the identity matrix, the ALP does not mix with the \(\eta^\prime\) at the
leading order (LO).

To derive the LO modifications to the \(K\pi\) FFs, we match the Lagrangian in
\cref{eq:LagQ} to the chiral Lagrangian written in terms of the exponential
representation of the mesons, \(U_\pi\), which transforms as a bi-fundamental
under $SU(3)_L \times SU(3)_R$:
\begin{align}
	\label{eq:eq2.15}
	U_\pi \ \equiv \ \exp\left( \nicefrac{2i\pi^a t^a}{f_\pi}  \right)\ \xrightarrow[L\times R]{}\
	L \: U_\pi \: R^\dag,
\end{align}
where \(L\) and \(R\) represent the \(SU(3)_L\) and  the \(SU(3)_R\)
transformations, respectively. In terms of \(U_\pi\), the chiral Lagrangian to
LO in the chiral expansion is:
\begin{align}
	\label{eq:eq2.16}
    \begin{split}
	\mathcal{L} \ &\supset \   \frac{f_\pi^2}{4} \:  \text{Tr}
	\Big[\left| \partial_\mu U_\pi - i (L_\mu U_\pi - U_\pi R_\mu) \right|^2 \Big] \\
	&\quad + \frac{\lambda f_\pi^2}{2} \: \text{Tr} \Big[  \overline{M}   U^\dagger_\pi  \Big] +\text{h.c.}\;,\\
    \mathrm{with}\ L_\mu\ &=\ L_\mu^\mathrm{SM} + \frac{\partial^\mu a}{f_a}
    \gamma^\mu C^{8}_{L} t_{8}^{L}\;,\quad R_\mu\ =\ R_\mu^\mathrm{SM} \\
            \overline{M}\ &=\ iC_{LR}^8\frac{a}{f_a}Mt^8\;,
   \end{split}
\end{align}
where \(L_\mu^\mathrm{SM}\) and \(R_\mu^\mathrm{SM}\) have the standard expressions~\cite{Georgi:1986df}.

Having matched the UV Lagrangian to the \(\chi\)PT in the IR, we can write the 
overlaps in \cref{eq:mix} in terms of the Wilson coefficients as
\begin{subequations}
    \label{eq:mix_coeff}
\begin{align}
        C_{a\pi} &= \phi_{a\pi} - \frac{C_L^8}{2} \epsilon\;;\;
        &C_{a\eta} = - \phi_{a\eta} + \frac{C_L^8}{2}\;;\\
        C_{\pi a} &= -\phi_{a\pi} + \phi_{a\eta} \epsilon\;;\;
        &C_{\eta a} = \phi_{a\pi} \epsilon + \phi_{a\eta}\;,
\end{align}
\end{subequations}
where, the \(\pi\)-\(\eta\) mixing angle, \(\epsilon\), is $1.16(13)\times
10^{-2}$~\cite{ParticleDataGroup:2024cfk} and we expand up to first order in it.
The quantities \(\phi_{a\pi}\) and \(\phi_{a\eta}\) satisfy the consistency
relations:
\begin{align}
    \label{eq:ang_def}
    \phi_{a\pi} &\simeq \frac{C_{LR}^8}{6} \frac{m_\sigma\, (B\, \epsilon 
            -\sqrt{3}m_\Delta)}{M_a^2-M_\pi^2} 
            + \frac{ C_L^8 \epsilon}{2} \frac{M_a^2}{M_a^2-M_\pi^2}\;;\notag\\
    \phi_{a\eta} &\simeq \frac{C_{LR}^8}{6} \frac{m_\sigma B}{
        M_a^2-M_\eta^2} + \frac{ C_L^8}{2} \frac{M_a^2}{M_a^2-M_\eta^2}\;.
\end{align}
In this equation, \(M_a\), \(M_\pi\), \(M_\eta\) are the pole masses of \(a\),
\(\pi^0\), and \(\eta\), respectively. We have kept terms to first order in
\(\xi\) and \(\epsilon\), and we have defined  \( m_\Delta\ =\ m_u-m_d;\
m_\sigma\ =\ m_u+m_d+4m_s\).
Clearly, the overlaps are ill-defined at the \(\pi^0\) and \(\eta\) poles and
hence our results are valid at points away from the resonances\footnote{We take the window to be a conservative 20\% of the meson mass. Note, even if
the ALP couples, for example, only to leptons at leading order, there will be
\(a\)-\(\pi\) mixing at one-loop level, which will be enhanced to non-perturbative
values at the poles. Hence, results at the \(\pi^0\) pole and the \(\eta\) pole
will always be ill-defined.}.

We can schematically divide the newly obtained chiral Lagrangian, the \(\mathcal{L}_{a\chi\mathrm{PT}}\), into two pieces, 
\begin{align}
    \mathcal{L}_{a\chi\mathrm{PT}}\ &= \mathcal{L}_{a\mathrm{SM}} + \mathcal{L}_{a\mathrm{NP}}\,,
\end{align}
where \(\mathcal{L}_{a\mathrm{SM}}\) contains operators made out of only SM-like
fields, and \(\mathcal{L}_{a\mathrm{NP}}\) has operators with an explicit \(a\)
degree of freedom. For our analyses, we concentrate exclusively on the
\(\mathcal{L}_{a\mathrm{SM}}\) part of the Lagrangian. Hence, the results are
sensitive only to the modifications to the SM-like interactions and insensitive
to the branching fractions of the ALP. This makes the analyses robust against
any coupling of the ALP to leptons or to a dark sector since the ALP is not
present in the final or the initial state, neither as an intermediate state. We can also combine multiple SM meson
decay modes, which depend on the same form factors, to obtain a tighter
constraint on the coupling, which is the spirit of the paper. 

In this redefined \(\chi\)PT, in the presence of the ALP, the \(\langle
K^+|\overline{s}\gamma^\mu u|\pi^0\rangle\) FFs are modified. In the absence of
first-principles calculations of the SM contributions to the FFs for the entire
kinematic range of the corresponding decay rates, the NP modifications cannot be
probed independently. Therefore, when data are used to extract FF parameters,
the modifications made by NP, are \emph{fitted away}. This issue can be
circumvented if we consider SM calculations of FFs on the lattice, or, if we
extract the SM FFs from processes that are not modified by the NP in question.

For this work, we use lattice determination of the \(\langle
K^+|\overline{s}\gamma^\mu u|\pi^0\rangle\) FF for the truncated kinematic
region corresponding to \(K^+\to \pi^0\ell^+\nu\)~\cite{Carrasco:2016kpy}  and
we use BELLE data corresponding to \(\langle K^0|\overline{s}\gamma^\mu
u|\pi^+\rangle\) to determine the form factor over the full kinematic range of
\(\tau^-\to \pi^0 K^-\nu\)~\cite{Belle:2007goc, Boito:2010me}, after restriction
to the lattice range~\cite{Carrasco:2016kpy}. We note, at the order we are
working, the \(\langle K^0|\overline{s}\gamma^\mu u|\pi^+\rangle\) FF is
unaffected by the NP interactions considered, as the corresponding FCNC
operators are not switched on. The approximate isospin symmetry in the SM
ensures that the \(K^+\pi^0\) and \(K^0\pi^+\) FFs are related. Hence, from the
FFs extracted from \(\tau^-\to K_s\pi^-\nu_t\) by Belle, we can compute the SM
expectations for the extended kinematic region of the \(K^+\pi^0\) FF
relevant for \(\tau^-\to \pi^0 K^-\nu\) decays. 

The bin-by-bin residuals in the experimental data for \(K^+\to \pi^0\ell^+\nu\)
and that for \(\tau^-\to \pi^0 K^-\nu\), once combined, determine the allowed
regions for the \(a\)-\(\pi\) and \(a\)-\(\eta\) overlaps. 
Needless to say, this methodology can be generalized to various isospin-related
decay modes, to find complementary bounds on the parameter space.


\emph{The modification to the form-factors:} The \(a\)-\(\pi\) and the
\(a\)-\(\eta\) overlaps modify the couplings of the \(\pi^0\) and the \(\eta\)
in the \(\mathcal{L}_{a\mathrm{SM}}\) part of the chiral Lagrangian. The
amplitudes for the hadronic decays of the \(\tau\) and the semi-leptonic decays
of the \(K\), both have the structure: 
\begin{align}
    \mathcal{A}\!&=\! G_F C^\prime V_{\bar{s}u}\!\left[\tilde{f}_+^{\!K^+\!\pi^0}\!(p^2) 
        Q_\mu\! + \! \tilde{f}_-^{\!K^+\!\pi^0}(p^2) q_\mu\right]\!\!\bar{u}_\nu\!
        \gamma^\mu\! P_L\! v_{\ell}.
\end{align}
Here, \(\tilde{f}_+^{K^+\pi^0}(p^2)\) and \(\tilde{f}_-^{K^+\pi^0}(p^2)\) are
the vector and the axial-vector form-factors. The momentum transfer \(p^2\) is
given by the \(t\) channel \(q_\mu=p_\mu^+-p_\mu^0\) for the \(K\) decay and the
\(s\) channel \(Q_\mu=p_\mu^++p_\mu^0\) for the \(\tau\) decay. The
\(\widetilde{~}\) indicates that the amplitudes are in the mass basis. The
pre-factors \(G_F\) and \(V_{\bar{s}u}\) are the Fermi constant and the relevant
CKM element, respectively. The \(C^\prime\) encapsulates short- and
long-distance EM corrections and electroweak corrections~\cite{Sirlin:1981ie,
Marciano:1988vm, Braaten:1990ef, Erler:2002mv,
FlaviaNetWorkingGrouponKaonDecays:2008hpm, Antonelli:2013usa,
Flores-Baez:2013eba, Pich:2013lsa,  Moulson:2017ive}. For \(V_{\bar{s}u}\), we
use the value obtained from the ratio of the \(K^+\to\mu\nu\) width to the
\(\pi^+\to\mu\nu\) width~\cite{Marciano:2004uf,ParticleDataGroup:2024cfk}.

\renewcommand*{\arraystretch}{1.4}
\begin{table*}[t]
    \centering
    \resizebox{\textwidth}{!}{
    \begin{tabular}{|c|c|c|c|c|}
\hline
Parameters &$\tau \to K\pi^0\nu_\tau$ Dist. + Width &$\tau \to K\pi^0\nu_\tau$~{\rm total} + $K_{\ell 3} $ &$\tau \to K\pi^0\nu_\tau$~({\rm Belle~Proj.}) &$\tau \to K\pi^0\nu_\tau$~({\rm Belle-II~Proj.})\\
\hline
$\xi^2\alpha$ & $ -0.046^{+0.028}_{-0.016}$ & $ -0.0090\pm 0.0079$ & $\pm 0.012$ & $\pm 0.0012$\\
$\xi^2\beta$ & $ 0.007^{+0.046}_{-0.027}$ & $ 0.0034\pm 0.0094$ & $\pm 0.011$ & $\pm 0.00090$\\
\hline
\end{tabular}
    }
    \caption{Marginalised constraints and future projections at $1\sigma$ on the
    form factor modification parameters \(\xi^2\alpha\) and \(\xi^2\beta\). For
    the constraints, we use \(\tau^-\to K^-\pi^0\nu\) differential data from
    BaBar and \(\tau\) width along with the NA48/2 data for \(K^+\to \pi^0
    \ell^+\nu\) and \(K^+\) width. For projections, we use \(\tau^-\to
    K^-\pi^0\nu\) estimations for Belle data and Belle II.}
    \label{tab:2d_v2}
\end{table*}

\begin{figure*}[t]
    \centering
    \includegraphics[width=0.35\textwidth]{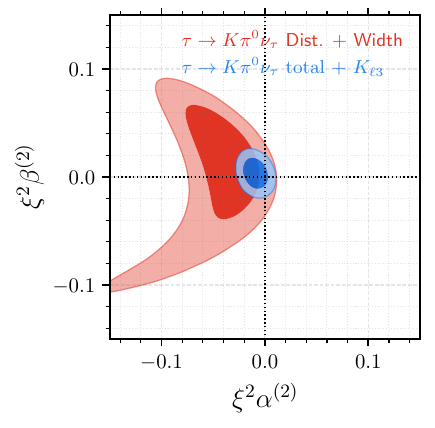}\quad
    \includegraphics[width=0.35\textwidth]{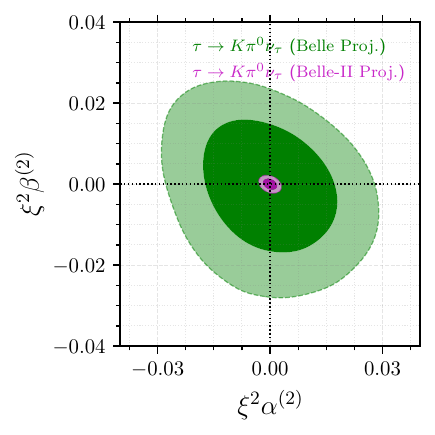}
    \caption{Constraints on form factor modifications $\xi^2\alpha$ and $
    \xi^2\beta$ is shown in the left plot. The shaded \textcolor{red}{red} regions show
    constraint at $68\%$ and $95\%$ C.L. from \(\tau^-\to K^-\pi^0\nu\)
    differential distribution and total decay width measurement. The
    \textcolor{blue}{blue} contours shows the combined constraint after
    including data from  \(K^+\to\pi^0\ell^+\nu_\ell (K_{\ell 3})\) differential
    rate and width measurement. The right plot shows the projection for Belle
    (green) and Belle-II (magenta) measurement of the same \(\tau^-\to
    K^-\pi^0\nu_\tau\) channel.} 
    \label{fig:2d_v1}
\end{figure*}

The modifications sourced by the ALP to the \(\tilde{f}_\pm^{K^+\pi^0}(0)\) are
two-fold. The first is due to the redefinition of the pion after \(a\)-\(\pi\)
mixing. The second is due to the ALP interactions in the flavor basis
getting `rotated' to generate additional pion interactions in the mass basis. We
parametrize the modified FFs at \(p^2=0\) as: 
\begin{align}
    \label{eq:fplus_mod}
    \begin{split}
    &\tilde{f}_{+(-)}^{k^+\pi^0}(0) = {f}_{+(-)}^{k^+\pi^0}(0) + \xi^2\alpha(\beta)\,;\\
    &\beta\! =\! -\frac{\sqrt{3}}{2}C_L^8 \phi_{a\pi};\,
    \alpha\! =\! \sqrt{3}\phi_{a\pi}\phi_{a\eta} - \frac{1}{2}\phi_{a\pi}^2,
    \end{split}
\end{align}
\({f}_{+(-)}^{k^+\pi^0}(0)\) is the FF before \(a-\pi\) mixing, that is, the SM 
expectation.

\(K\) and \(\tau\) decays depend on different kinematic ranges of the $K\pi$
form factor due to the difference in available phase space. For the decay of the
\(K^\pm\), we can expand the FFs in \(t=q^2=(q_{K}-q_{\pi})^2\) in a Taylor
series~\cite{Cirigliano:2001mk}:
\begin{align}
    &\tilde{f}_{+(-)}^{K\pi}(t)=\left( 
        {f}_{+(-)}^{K\pi}(0) + \xi^2\alpha(\beta)\right)
        F_1\!\left(t;\lambda^{+(-)}_{K\pi}, \lambda^{\prime +(-)}_{K\pi}\right)\,; \notag\\
    &\mathrm{with},\ F_1(t;\lambda_1, \lambda_2) 
    = 1+\lambda_1\frac{t}{M_{\pi^+}^2}
            +  \lambda_2\frac{t^2}{M_{\pi^+}^4}\;.
\end{align}
Here, \(\lambda^{(\prime)+(-)}\) are the slope factors determining the variation
of the FF with \(p^2\). The parametrisation is the same for \(K^+\pi^0\) and
\(K^0\pi^+\) FFs. Note that we do not consider the sub-leading NP corrections
to the slope factors, and we extract their values from
Ref.~\cite{Bernard:2009zm}. For the FF values at \(q^2=0\), we use Lattice
determinations of the FF~\cite{Carrasco:2016kpy}.  

For the decays of the \(\tau\), the kinematically allowed region is
\(M_K^2+M_\pi^2 < s=Q^2< M_\tau^2\), contaminated by resonances. To tackle
these, we resort to the thrice-subtracted dispersive vector FF as given by
\cite{Guerrero:1997ku, Pich:2001pj, Boito:2010me}:
\begin{align}
\label{eq: vff}
	  &\tilde{f}_{+}^{K\pi}(t)= \left(\tilde{f}_{+}^{K\pi}(0) 
        + \xi^2\alpha\right) F_2(s; \alpha_1, \alpha_2)\,; \notag\\
    &\mathrm{with,}\  F_2(s; \alpha_1, \alpha_2) =\exp\left[\alpha_1 \,\frac{s}{M^2_{\pi^+}} + \frac{\alpha_2}
    {2}\,\left(\frac{s}{M^2_{\pi^+}}\right)^2 \right. \quad \notag\\
    &\quad + \left. \frac{s^3}{\pi}\int_{(m_K+m_\pi)^2}^{s_{\rm cut}}ds^\prime \frac{\delta_1(s^\prime)}{(s^\prime)^3(s^\prime - s -i\epsilon)}\right].
\end{align}
We extract the elastic and inelastic components of the phase \(\delta_1(s)\)
from resonance chiral perturbation theory results, following Refs.
\cite{Jamin:2006tk,Jamin:2008qg, Boito:2008fq, Gasser:1984ux}.

The axial-vector FF is obtained using the relation, 
\begin{equation}
    \label{eq:fm_f0}
    \tilde{f}_-^{K\pi}(s) = \frac{\Delta_{K\pi}}{s}\left(\tilde{f}_0^{K\pi}(s) - \tilde{f}_+^{K\pi}(s)\right)\,,
\end{equation}
where \(K\pi\) can be both \(K^+\pi^0\) and \(K^0\pi^+\). For the scalar FF, \(\tilde{f}_0^{K^+\pi^0}(s) \), expected to give
important contributions to the low-energy region ($\lesssim 0.8$ GeV) of the
$\tkspn$ spectrum \cite{Jamin:2008qg, Boito:2008fq}, we use the
coupled-channel dispersive parametrisation~\cite{Jamin:2001zq}. In our numerical
analysis, we use the updated fitted parameters for $s_{\rm cut} = 4~{\rm
GeV}^2$ \cite{Jamin:2008qg} with the corresponding uncertainties from
Ref.~\cite{Boito:2008fq}.

\emph{Numerical analysis:}
The differential decay rates for $\tau^- \to K^-\pi^0 \nu$ and \(K^+\to
\pi^0\ell^+\nu\) as a function of the hadronic invariant mass squared, are
computed in Ref.~\cite{Pich:2013lsa} and Refs.~\cite{Bandyopadhyay:2021wbb}
respectively. We use those results and \cref{eq:isospin_corr} and
\cref{eq:fplus_mod}, to compute the differential rate and the total width
of \(\tau^-\to K^-\pi^0\nu\) and \(K^+\to \pi^0\ell^+\nu\) to constrain the
effective NP parameters $\xi^2\alpha$ and
$\xi^2\beta$. Using the map provided in \cref{eq:fplus_mod},
we can relate these modifications to the Wilson coefficients.

\begin{figure*}[t]
    \centering
    \includegraphics[width=0.47\textwidth]{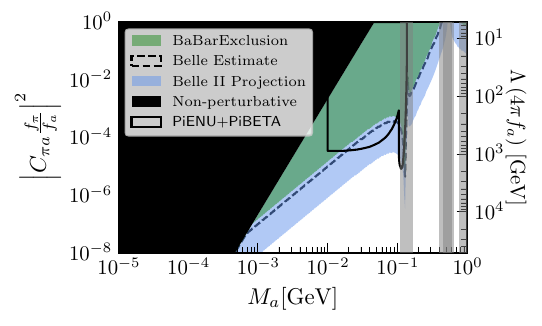}
    \includegraphics[width=0.47\textwidth]{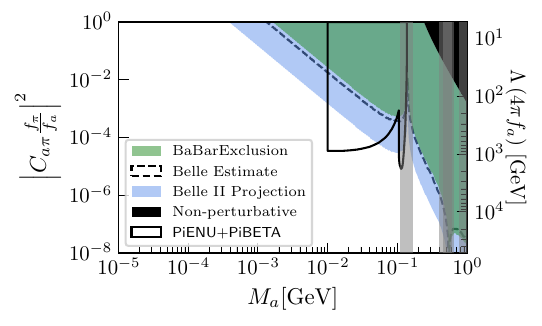}
    \includegraphics[width=0.47\textwidth]{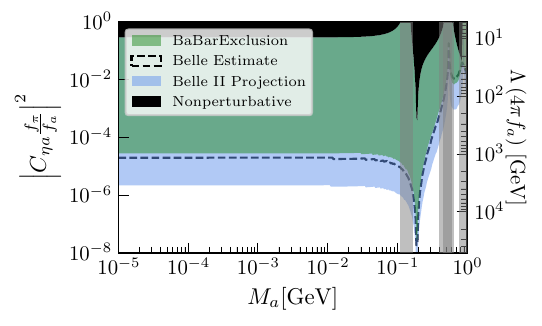}
    \includegraphics[width=0.47\textwidth]{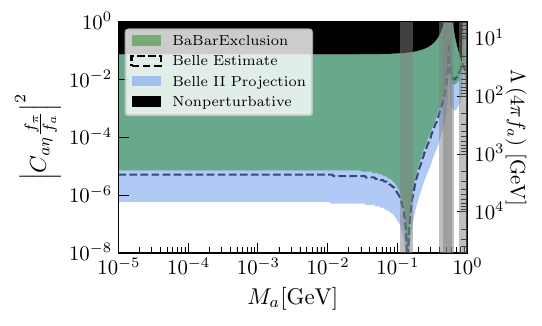}
    \caption{Constraints on the overlaps between the ALP and the \(\pi^0\) and
    \(\eta\) mesons. We have the \(\pi\)-\(a\) overlap (top left), the
    \(a\)-\(\pi\) overlap (top right), the \(\eta\)-\(a\) overlap (bottom left)
    and the \(a\)-\(\eta\) overlap (bottom right). The green regions indicate
    the parameter space that we can rule out from BaBar data, the black dashed
    boundaries indicate the regions we estimate would be ruled out from a Belle
    analysis, and the blue shaded regions indicate the areas that we project to
    be accessible by Belle II. In solid black, we have the regions in the
    mass-overlap plane that are not accessible by perturbative values of the
    Wilson coefficients. We have grayed out narrow regions around the neutral meson masses (\(\pi^0\), \(K^0\), \(\eta\), \(\eta^\prime\)), at which the 
    overlaps blow up. We have also indicated by the solid black boundary, existing bounds on \(a\)-\(\pi^0\) mixing from the PiENU and PiBETA experiments \cite{Altmannshofer:2019yji}.}
    \label{fig:2d_v2}
\end{figure*}

For the \(K^+\to\pi^0\ell^+\nu\) decay, we consider the differential decay rate
recorded by the NA48/2 collaboration~\cite{NA482:2018rgv}. For \(\tau^-\to
K^-\pi^0\nu\) we use the decay rate and total decay width measurement from the
BaBar collaboration~\cite{BaBar:2007yir,ParticleDataGroup:2024cfk}. For the SM
computations of the FFs, we extract the FF for \(\tau^-\to\pi^-K^0\nu\) from the
Belle data in Ref.~\cite{Belle:2007goc} and compute our desired FF for
\(\tau^-\to\pi^0K^-\nu\) by using the relation~\cite{Antonelli:2013usa}:
\begin{equation}
    \label{eq:isospin_corr}
    \frac{\tilde{f}_{+(0)}^{K^{+}\pi^{0}}(q^2)}{\tilde{f}_{+(0)}^{\bar{K^{0}}\pi^{-}}
    (q^2)} = (1+\sqrt{3}\epsilon)\left(1+\tilde{g}{m_K^2 \over (4\pi
    f_\pi)^2}{q^2 \over m_{K_\ast}^2}\epsilon\right)
\end{equation}
The first term in Eq.~\cref{eq:isospin_corr} originates from $\pi^0$-$\eta$
mixing. The momentum-dependent correction stems from isospin-breaking effects
in the coupling of $K^-_\ast$ to $K\pi$. We employed the same isospin correction
to the \(\tilde{f}_0^{K\pi}(s)\) and used Eq.~\eqref{eq:fm_f0} to derive
\(\tilde{f}_{-}^{K^{+}\pi^{0}}(q^2)\).  

An advantage of constraining new physics via FF modification is that we can
combine various channels that are sensitive to the same FFs. We perform an MCMC
analysis to constrain the NP parameters \(\xi^2\alpha\) and \(\xi^2\beta\) using
the binned \(K^+\to \pi^0\ell^+\nu\) decay distribution from
BaBar~\cite{BaBar:2007yir} and the total decay
width~\cite{ParticleDataGroup:2024cfk}. We use flat priors on \(\xi^2\alpha\)
and \(\xi^2\beta\) for all the MCMC analyses. To derive a tighter constraint, we
also combined the NA48/2 measurement of the \(K_{\ell 3}\) decays
distribution~\cite{NA482:2018rgv} and the total decay
width~\cite{ParticleDataGroup:2024cfk}, following our previous analysis in
Ref.~\cite{Bandyopadhyay:2021wbb}.

The red contours in \cref{fig:2d_v1} show the $68\%$ and $95\%$ constraints on
the NP parameters, obtained by analyzing BaBar \(K^+\to \pi^0\ell^+\nu\)
data~\cite{BaBar:2007yir,ParticleDataGroup:2024cfk}, while the blue contours
show the constraint when the BaBar and NA48/2 $K_{\ell 3}$ data are combined.
The SM prediction (\(\xi^2\alpha = \xi^2\beta = 0\)) is consistent with the limits
thus derived. In our analysis, we properly take into account the theoretical
uncertainty of the FF parameters, assuming a normal distribution. We added the
theoretical uncertainty in quadrature to the experimental uncertainty coming
from the error bars of the binned BaBar and NA48/2 data. In \cref{tab:2d_v2}, we
document the marginalized \(1\sigma\) constraint of the individual parameters.

BaBar's sister experiment, Belle, has achieved twice the integrated luminosity
compared to the former. But, the similar analysis for the \(\tau^-\to
K^-\pi^0\nu_\tau\) as in Ref.~\cite{BaBar:2007yir} has not been performed 
with Belle data. However, Belle has performed a measurement for the iso-spin
conjugate channel \(\tau^-\to K_S\pi^-\nu_\tau\)~\cite{Belle:2007goc}. 
To obtain the Belle expectation, we made the reasonable assumption that the
\(\tau^-\to K^-\pi^0\nu_\tau\) differential distribution can be measured with a
similar fractional per-bin error as of  \(\tau^-\to K_S\pi^-\nu_\tau\)
differential distribution. Thus, we assume that the fractional errorbar of the
$i$th bin for the \(\tau^-\to K^-\pi^0\nu_\tau\) distribution will be $\Delta
N_i/N_i$ where $N_i$ and $\Delta N_i$ are the mean and standard deviation of the
measured events for \(\tau^-\to K_S\pi^-\nu_\tau\). Fig.~\ref{fig:2d_v1} right
panel shows the constraining power from that projection in the using green
contours. The marginalized $1\sigma$ constraining power (errorbar) for each new
physics parameter is shown in Table.~\ref{tab:2d_v2}. Note that the improvement
of the Belle constraining power is within a factor of $2$ compared to the
BaBar+NA48/2 analysis.

Finally, we performed a projection for Belle II, which is expected to collect
data at an integrated luminosity of $50~{\rm ab}^{-1}$. We used the Belle data
to estimate the differential distribution for \(\tau^-\to K^-\pi^0\nu_\tau\). We
then scale the errorbar per bin by the square root of the luminosity ratio
between Belle and Belle-II, i.e, by a factor $\sqrt{50/0.669} \approx 9$. We
also assumed that the theoretical uncertainty coming from the FF measurements is
also reduced by the same factor. This is an aggressive assumption that implies
that some systematic uncertainties will also scale similarly to statistical
uncertainties, which is often not the case. The projection can be viewed as the
best-case scenario, which showcases the maximum potential of Belle II to probe
this kind of new physics. The right panel of Fig.~\ref{fig:2d_v1} shows the
constraining power of Belle II in magenta. Table.~\ref{tab:2d_v2} shows the
marginalized $1\sigma$ constraining power on individual parameters. The
constraining power of Belle-II projection is roughly one order of magnitude
stronger than Belle, which is in line with the scaling of total luminosity.
Note, as for Belle and Belle II, we only have projections; we do not combine the
\(K_{\ell_3}\) data with them.

The bounds obtained on \(\xi^2 \alpha\) and \(\xi^2 \beta\) can be mapped back
to the Wilson coefficients by numerically inverting the functions in
\cref{eq:fplus_mod}. The bounds on the Wilson coefficients end up constraining
the overlaps through \cref{eq:mix_coeff,eq:ang_def}. We plot the 95\% C.L.
limits on  \(C_{\pi a} \xi^2\), \(C_{a\pi} \xi^2\), \(C_{\eta a} \xi^2\), and
\(C_{a\eta} \xi^2\) in the upper-left, upper-right, lower-left, and lower-right
panels of \cref{fig:2d_v2}, respectively. We keep one of the
two Wilson coefficients at zero while deriving the bounds. For \(C_{a\pi}\) and
\(C_{\pi a}\) we keep \(C_{LR}^8=0\), while for \(C_{a\eta}\) and \(C_{\eta
a}\), we keep \(C_L^8=0\). 

In green, we plot the limits from the BaBar experiment, using the black dashed
line we indicate the region expected to be ruled out by the Belle data, and in blue, we
give the Belle II projections. In black, we indicate the regions that are
inaccessible via perturbative values of the Wilson coefficients \(C<4\pi\). For
the \(a\)-\(\pi\) overlaps, we also show the bounds obtained from the PiENU and
PiBETA experiments from Ref. \cite{Altmannshofer:2019yji}. The  PiENU and PiBETA
bounds are confined to a specific window in the ALP mass, owing to experimental
constraints. However, as our bounds are obtained from processes where the
ALP doesn't take part, they extend over all ALP masses. By considering the
Wilson coefficients to be one, we translate the limits to the effective scale,
\(\Lambda=4\pi f_a\), associated with integrating out the UV physics giving rise
to the ALP. We find that the effective scale is bounded till multi-TeV ranges
from all the limits, for regions of the ALP mass. Overall, the bounds on the
\(a\)-\(\eta\) overlaps are stronger than the \(a\)-\(\pi\) overlaps. 

We stress that Belle II data can give the strongest bound as yet on the
\(\pi\)-\(a\) overlap, while the BaBar data already constrain the \(\eta\)-\(a\)
and \(a\)-\(\eta\) bounds to \(\mathcal{O}\)(TeV). To our knowledge, analogous
bounds on the \(a\)-\(\eta\) overlaps do not exist in the literature, while
separate bounds that treat the formally different objects \(C_{a\pi}\) and
\(C_{\pi a}\) differently are also absent.

\emph{Conclusion:}
Even for a simplistic ALP-quark Lagrangian, the notion of ALP-meson mixing is
nuanced, with a general linear transformation connecting the flavor and mass
bases. As a result, ALP-meson and meson-ALP overlaps are different objects. We
take this into account while constraining the ALP-\(\pi^0\), \(\pi^0\)-ALP,
ALP-\(\eta\), and \(\eta\)-ALP overlaps against ALP mass. We show that, in
terms of the characteristic scale, \(\Lambda =4\pi f_a\), the overlaps are
already strongly excluded---multi-TeV---for substantial ranges of the ALP mass
from BaBar and NA48/2 data. We also estimate the reach of a Belle analysis
and give projections for Belle II. Bounds that treat ALP-\(\pi^0\) and
\(\pi^0\)-ALP overlaps separately are absent in the literature. So too are
bounds on ALP-\(\eta\) and \(\eta\)-ALP mixing. The proposed REDTOP
experiment~\cite{Zielinski:2025wfa} has the potential to perform precision
\(\eta\) physics that can be used to experimentally probe the ALP-\(\eta\) and
\(\eta\)-ALP overlaps. A key feature of our approach is that our analysis does
not depend on processes with the ALP in the external or internal lines. Hence,
all the limits we derive are independent of the branching ratios of the ALP, and
consequently of many model-specific assumptions about ALP decays. We note that
our analysis is designed around a restricted set of UV operators, specifically,
\(t_8\) isospin-breaking interactions, and works at leading order in the chiral
expansion. A natural extension would be to marginalize over a broader set of
Wilson coefficients and to assess the impact of next-to-leading order
corrections. Finally, we urge the Belle II Collaboration to analyze the
differential distributions pertaining to the \(\tau^-\to K^-\pi^0\nu\) and
\(\tau^-\to K_s\pi^-\nu\) channels, to look for possible tensions between the
two, that might lead to the hint of such a particle. This will be a relevant
addition to their planned suite of analyses to look for new physics in \(\tau\)
decay channels.  

\section*{Acknowledgement}
We thank Girish Kumar for multiple discussions and clarifications on hadronic
decays of $\tau$ and for reviewing an advanced version of the paper. We also
thank Pablo Roig for providing the data files related to the scalar form-factor
$(F_0(s))$. We thank D.~Epifanov and G.~Mohanty of the Belle Collaboration for
providing us with the $\tau\to K_s\pi\nu$ spectrum data file. We also thank  Dmitry
Madigozhin for providing us with the source of the NA48/2 dataset. SG
acknowledges support from the National Science Foundation (NSF) under Grant No.
PHY2413016. This work used the high-performance computing service at the
University of Notre Dame, managed by the Center for Research Computing (CRC)
(\href{https://crc.nd.edu}{https://crc.nd.edu}). We use
\texttt{emcee}~\cite{emcee} to run the MCMC chains and
\texttt{GetDist}~\cite{Lewis:2019xzd} for deriving the constraints and plotting.
For general analyses and visualizations, we have used \texttt{Python 3.13.7} and its
scientific stack of \texttt{Numpy}~\cite{harris2020array},
\texttt{SciPy}~\cite{2020scipy-nmeth},and \texttt{Matplotlib}~\cite{Hunter:2007}. 

\bibliography{refs}

@article{Pich:2013lsa,
    author = "Pich, Antonio",
    title = "{Precision Tau Physics}",
    eprint = "1310.7922",
    archivePrefix = "arXiv",
    primaryClass = "hep-ph",
    reportNumber = "IFIC-13-79, FTUV-13-1029, FTUV-13-1028",
    doi = "10.1016/j.ppnp.2013.11.002",
    journal = "Prog. Part. Nucl. Phys.",
    volume = "75",
    pages = "41--85",
    year = "2014"
}

@article{Boito:2008fq,
    author = "Boito, Diogo R. and Escribano, Rafel and Jamin, Matthias",
    title = "{K pi vector form-factor, dispersive constraints and tau ---\ensuremath{>} nu(tau) K pi decays}",
    eprint = "0807.4883",
    archivePrefix = "arXiv",
    primaryClass = "hep-ph",
    reportNumber = "UAB-FT-653",
    doi = "10.1140/epjc/s10052-008-0834-9",
    journal = "Eur. Phys. J. C",
    volume = "59",
    pages = "821--829",
    year = "2009"
}

@article{Boito:2010me,
    author = "Boito, D. R. and Escribano, R. and Jamin, M.",
    title = "{K $\pi$ vector form factor constrained by $\tau -> K\ pi \nu_\tau$ and $K_{l3}$ decays}",
    eprint = "1007.1858",
    archivePrefix = "arXiv",
    primaryClass = "hep-ph",
    reportNumber = "UAB-FT-682",
    doi = "10.1007/JHEP09(2010)031",
    journal = "JHEP",
    volume = "09",
    pages = "031",
    year = "2010"
}

@article{Jamin:2006tk,
    author = "Jamin, Matthias and Pich, Antonio and Portoles, Jorge",
    title = "{Spectral distribution for the decay tau ---\ensuremath{>} nu(tau) K pi}",
    eprint = "hep-ph/0605096",
    archivePrefix = "arXiv",
    reportNumber = "UAB-FT-601, IFIC-06-10, FTUV-06-0509",
    doi = "10.1016/j.physletb.2006.06.058",
    journal = "Phys. Lett. B",
    volume = "640",
    pages = "176--181",
    year = "2006"
}

@article{Jamin:2008qg,
    author = "Jamin, Matthias and Pich, Antonio and Portoles, Jorge",
    title = "{What can be learned from the Belle spectrum for the decay - tau- ---\ensuremath{>} nu(tau) K(S) pi-}",
    eprint = "0803.1786",
    archivePrefix = "arXiv",
    primaryClass = "hep-ph",
    reportNumber = "UAB-FT-642, IFIC-08-16, FTUV-08-0312",
    doi = "10.1016/j.physletb.2008.04.049",
    journal = "Phys. Lett. B",
    volume = "664",
    pages = "78--83",
    year = "2008"
}

@article{Gasser:1984ux,
    author = "Gasser, J. and Leutwyler, H.",
    title = "{Low-Energy Expansion of Meson Form-Factors}",
    reportNumber = "CERN-TH-3829/84",
    doi = "10.1016/0550-3213(85)90493-6",
    journal = "Nucl. Phys. B",
    volume = "250",
    pages = "517--538",
    year = "1985"
}

@article{Jamin:2001zq,
    author = "Jamin, Matthias and Oller, Jose Antonio and Pich, Antonio",
    title = "{Strangeness changing scalar form-factors}",
    eprint = "hep-ph/0110193",
    archivePrefix = "arXiv",
    reportNumber = "IFIC-01-26, FTUV-01-1015, HD-THEP-01-10, FZ-IKP-TH-01-16",
    doi = "10.1016/S0550-3213(01)00605-8",
    journal = "Nucl. Phys. B",
    volume = "622",
    pages = "279--308",
    year = "2002"
}

@article{Belle:2007goc,
    author = "Epifanov, D. and others",
    collaboration = "Belle",
    title = "{Study of tau- ---\ensuremath{>} K(S) pi- nu(tau) decay at Belle}",
    eprint = "0706.2231",
    archivePrefix = "arXiv",
    primaryClass = "hep-ex",
    reportNumber = "BELLE-PREPRINT-2007-28, KEK-PREPRINT-2007-17",
    doi = "10.1016/j.physletb.2007.08.045",
    journal = "Phys. Lett. B",
    volume = "654",
    pages = "65--73",
    year = "2007"
}

@article{Bandyopadhyay:2021wbb,
    author = "Bandyopadhyay, Triparno and Ghosh, Subhajit and Roy, Tuhin S.",
    title = "{ALP-Pions generalized}",
    eprint = "2112.13147",
    archivePrefix = "arXiv",
    primaryClass = "hep-ph",
    doi = "10.1103/PhysRevD.105.115039",
    journal = "Phys. Rev. D",
    volume = "105",
    number = "11",
    pages = "115039",
    year = "2022"
}

@article{Guerrero:1997ku,
    author = "Guerrero, Francisco and Pich, Antonio",
    title = "{Effective field theory description of the pion form-factor}",
    eprint = "hep-ph/9707347",
    archivePrefix = "arXiv",
    reportNumber = "FTUV-97-42, IFIC-97-42",
    doi = "10.1016/S0370-2693(97)01070-8",
    journal = "Phys. Lett. B",
    volume = "412",
    pages = "382--388",
    year = "1997"
}

@article{Pich:2001pj,
    author = "Pich, A. and Portoles, J.",
    title = "{The Vector form-factor of the pion from unitarity and analyticity: A Model independent approach}",
    eprint = "hep-ph/0101194",
    archivePrefix = "arXiv",
    reportNumber = "IFIC-01-02, FTUV-01-0117",
    doi = "10.1103/PhysRevD.63.093005",
    journal = "Phys. Rev. D",
    volume = "63",
    pages = "093005",
    year = "2001"
}

@article{Peccei:1977hh,
    author = "Peccei, R. D. and Quinn, Helen R.",
    title = "{CP Conservation in the Presence of Instantons}",
    reportNumber = "ITP-568-STANFORD",
    doi = "10.1103/PhysRevLett.38.1440",
    journal = "Phys. Rev. Lett.",
    volume = "38",
    pages = "1440--1443",
    year = "1977"
}

@article{Peccei:1977ur,
    author = "Peccei, R. D. and Quinn, Helen R.",
    title = "{Constraints Imposed by CP Conservation in the Presence of Instantons}",
    reportNumber = "ITP-572-STANFORD",
    doi = "10.1103/PhysRevD.16.1791",
    journal = "Phys. Rev. D",
    volume = "16",
    pages = "1791--1797",
    year = "1977"
}

@article{Weinberg:1977ma,
    author = "Weinberg, Steven",
    title = "{A New Light Boson?}",
    reportNumber = "HUTP-77/A074",
    doi = "10.1103/PhysRevLett.40.223",
    journal = "Phys. Rev. Lett.",
    volume = "40",
    pages = "223--226",
    year = "1978"
}

@article{Wilczek:1977pj,
    author = "Wilczek, Frank",
    title = "{Problem of Strong  $P$  and  $T$  Invariance in the Presence of Instantons}",
    reportNumber = "Print-77-0939 (COLUMBIA)",
    doi = "10.1103/PhysRevLett.40.279",
    journal = "Phys. Rev. Lett.",
    volume = "40",
    pages = "279--282",
    year = "1978"
}

@article{tHooft:1976rip,
    author = "'t Hooft, Gerard",
    editor = "Shifman, Mikhail A.",
    title = "{Symmetry Breaking Through Bell-Jackiw Anomalies}",
    reportNumber = "PRINT-76-0254 (HARVARD)",
    doi = "10.1103/PhysRevLett.37.8",
    journal = "Phys. Rev. Lett.",
    volume = "37",
    pages = "8--11",
    year = "1976"
}

@article{Preskill:1982cy,
    author = "Preskill, John and Wise, Mark B. and Wilczek, Frank",
    editor = "Srednicki, M. A.",
    title = "{Cosmology of the Invisible Axion}",
    reportNumber = "HUTP-82-A048, NSF-ITP-82-103",
    doi = "10.1016/0370-2693(83)90637-8",
    journal = "Phys. Lett. B",
    volume = "120",
    pages = "127--132",
    year = "1983"
}

@article{Dine:1982ah,
    author = "Dine, Michael and Fischler, Willy",
    editor = "Srednicki, M. A.",
    title = "{The Not So Harmless Axion}",
    reportNumber = "UPR-0201T",
    doi = "10.1016/0370-2693(83)90639-1",
    journal = "Phys. Lett. B",
    volume = "120",
    pages = "137--141",
    year = "1983"
}

@article{Abbott:1982af,
    author = "Abbott, L. F. and Sikivie, P.",
    editor = "Srednicki, M. A.",
    title = "{A Cosmological Bound on the Invisible Axion}",
    reportNumber = "PRINT-82-0695 (BRANDEIS)",
    doi = "10.1016/0370-2693(83)90638-X",
    journal = "Phys. Lett. B",
    volume = "120",
    pages = "133--136",
    year = "1983"
}

@article{Graham:2015cka,
    author = "Graham, Peter W. and Kaplan, David E. and Rajendran, Surjeet",
    title = "{Cosmological Relaxation of the Electroweak Scale}",
    eprint = "1504.07551",
    archivePrefix = "arXiv",
    primaryClass = "hep-ph",
    doi = "10.1103/PhysRevLett.115.221801",
    journal = "Phys. Rev. Lett.",
    volume = "115",
    number = "22",
    pages = "221801",
    year = "2015"
}

@article{Hook:2016mqo,
    author = "Hook, Anson and Marques-Tavares, Gustavo",
    title = "{Relaxation from particle production}",
    eprint = "1607.01786",
    archivePrefix = "arXiv",
    primaryClass = "hep-ph",
    doi = "10.1007/JHEP12(2016)101",
    journal = "JHEP",
    volume = "12",
    pages = "101",
    year = "2016"
}

@article{Trifinopoulos:2022tfx,
    author = "Trifinopoulos, Sokratis and Vanvlasselaer, Miguel",
    title = "{Attracting the electroweak scale to a tachyonic trap}",
    eprint = "2210.13484",
    archivePrefix = "arXiv",
    primaryClass = "hep-ph",
    reportNumber = "MIT-CTP/5490",
    doi = "10.1103/PhysRevD.107.L071701",
    journal = "Phys. Rev. D",
    volume = "107",
    number = "7",
    pages = "L071701",
    year = "2023"
}

@article{Co:2019wyp,
    author = "Co, Raymond T. and Harigaya, Keisuke",
    title = "{Axiogenesis}",
    eprint = "1910.02080",
    archivePrefix = "arXiv",
    primaryClass = "hep-ph",
    reportNumber = "LCTP-19-27",
    doi = "10.1103/PhysRevLett.124.111602",
    journal = "Phys. Rev. Lett.",
    volume = "124",
    number = "11",
    pages = "111602",
    year = "2020"
}

@article{Chakraborty:2021fkp,
    author = "Chakraborty, Sabyasachi and Jung, Tae Hyun and Okui, Takemichi",
    title = "{Composite neutrinos and the QCD axion: Baryogenesis, dark matter, small Dirac neutrino masses, and vanishing neutron electric dipole moment}",
    eprint = "2108.04293",
    archivePrefix = "arXiv",
    primaryClass = "hep-ph",
    reportNumber = "KEK-TH-2341",
    doi = "10.1103/PhysRevD.105.015024",
    journal = "Phys. Rev. D",
    volume = "105",
    number = "1",
    pages = "015024",
    year = "2022"
}

@article{NA482:2018rgv,
    author = "Batley, John Richard and others",
    collaboration = "NA48/2",
    title = "{Measurement of the form factors of charged kaon semileptonic decays}",
    eprint = "1808.09041",
    archivePrefix = "arXiv",
    primaryClass = "hep-ex",
    reportNumber = "CERN-EP-2018-231",
    doi = "10.1007/JHEP10(2018)150",
    journal = "JHEP",
    volume = "10",
    pages = "150",
    year = "2018"
}

@software{madigozhin_dmitry_2019_3560600,
  author       = {Madigozhin Dmitry and
                  Shkarovskiy  Sergey and
                  NA48/2 collaboration},
  title        = {NA48/2 program and data for calculation of charged
                   kaon semileptonic form factors
                  },
  month        = dec,
  year         = 2019,
  publisher    = {Zenodo},
  doi          = {10.5281/zenodo.3560600},
  url          = {https://doi.org/10.5281/zenodo.3560600},
}

@article{BaBar:2007yir,
    author = "Aubert, Bernard and others",
    collaboration = "BaBar",
    title = "{Measurement of the $\tau^{-} \to K^{-} \pi^0 \nu_{tau}$ branching fraction}",
    eprint = "0707.2922",
    archivePrefix = "arXiv",
    primaryClass = "hep-ex",
    reportNumber = "SLAC-PUB-12681, BABAR-PUB-07-036",
    doi = "10.1103/PhysRevD.76.051104",
    journal = "Phys. Rev. D",
    volume = "76",
    pages = "051104",
    year = "2007"
}

@article{ParticleDataGroup:2024cfk,
    author = "Navas, S. and others",
    collaboration = "Particle Data Group",
    title = "{Review of particle physics}",
    doi = "10.1103/PhysRevD.110.030001",
    journal = "Phys. Rev. D",
    volume = "110",
    number = "3",
    pages = "030001",
    year = "2024"
}

@article{Altmannshofer:2019yji,
    author = "Altmannshofer, Wolfgang and Gori, Stefania and Robinson, Dean J.",
    title = "{Constraining axionlike particles from rare pion decays}",
    eprint = "1909.00005",
    archivePrefix = "arXiv",
    primaryClass = "hep-ph",
    doi = "10.1103/PhysRevD.101.075002",
    journal = "Phys. Rev. D",
    volume = "101",
    number = "7",
    pages = "075002",
    year = "2020"
}

@article{Carrasco:2016kpy,
    author = "Carrasco, N. and Lami, P. and Lubicz, V. and Riggio, L. and Simula, S. and Tarantino, C.",
    title = "{$K \to \pi$ semileptonic form factors with $N_f=2+1+1$ twisted mass fermions}",
    eprint = "1602.04113",
    archivePrefix = "arXiv",
    primaryClass = "hep-lat",
    reportNumber = "PREPRINT-RM3-TH-16-2",
    doi = "10.1103/PhysRevD.93.114512",
    journal = "Phys. Rev. D",
    volume = "93",
    number = "11",
    pages = "114512",
    year = "2016"
}

@article{Georgi:1986df,
    author = "Georgi, Howard and Kaplan, David B. and Randall, Lisa",
    title = "{Manifesting the Invisible Axion at Low-energies}",
    reportNumber = "HUTP-86/A004",
    doi = "10.1016/0370-2693(86)90688-X",
    journal = "Phys. Lett. B",
    volume = "169",
    pages = "73--78",
    year = "1986"
}

@article{Bauer:2020jbp,
    author = "Bauer, Martin and Neubert, Matthias and Renner, Sophie and Schnubel, Marvin and Thamm, Andrea",
    title = "{The Low-Energy Effective Theory of Axions and ALPs}",
    eprint = "2012.12272",
    archivePrefix = "arXiv",
    primaryClass = "hep-ph",
    reportNumber = "IPPP/20/69, MITP/20-070 SISSA 30/2020/FISI, ZH-TH-47/20",
    doi = "10.1007/JHEP04(2021)063",
    journal = "JHEP",
    volume = "04",
    pages = "063",
    year = "2021"
}

@article{Bauer:2021wjo,
    author = "Bauer, Martin and Neubert, Matthias and Renner, Sophie and Schnubel, Marvin and Thamm, Andrea",
    title = "{Consistent Treatment of Axions in the Weak Chiral Lagrangian}",
    eprint = "2102.13112",
    archivePrefix = "arXiv",
    primaryClass = "hep-ph",
    reportNumber = "IPPP/20-82, MITP/21-007, ZU-TH-01/21",
    doi = "10.1103/PhysRevLett.127.081803",
    journal = "Phys. Rev. Lett.",
    volume = "127",
    number = "8",
    pages = "081803",
    year = "2021"
}

@article{Bauer:2021mvw,
    author = "Bauer, Martin and Neubert, Matthias and Renner, Sophie and Schnubel, Marvin and Thamm, Andrea",
    title = "{Flavor probes of axion-like particles}",
    eprint = "2110.10698",
    archivePrefix = "arXiv",
    primaryClass = "hep-ph",
    reportNumber = "MITP/21-025, CERN-TH-2021-148, IPPP/21/37",
    doi = "10.1007/JHEP09(2022)056",
    journal = "JHEP",
    volume = "09",
    pages = "056",
    year = "2022"
}

@article{Sirlin:1981ie,
    author = "Sirlin, A.",
    title = "{Large m(W), m(Z) Behavior of the O(alpha) Corrections to Semileptonic Processes Mediated by W}",
    reportNumber = "NYU/TR8/81",
    doi = "10.1016/0550-3213(82)90303-0",
    journal = "Nucl. Phys. B",
    volume = "196",
    pages = "83--92",
    year = "1982"
}

@inproceedings{FlaviaNetWorkingGrouponKaonDecays:2008hpm,
    author = "Antonelli, Mario and others",
    collaboration = "FlaviaNet Working Group on Kaon Decays",
    title = "{Precision tests of the Standard Model with leptonic and semileptonic kaon decays}",
    booktitle = "{5th International Workshop on e+ e- Collisions from Phi to Psi}",
    eprint = "0801.1817",
    archivePrefix = "arXiv",
    primaryClass = "hep-ph",
    reportNumber = "FERMILAB-PUB-08-101-T",
    month = "1",
    year = "2008"
}

@article{Moulson:2017ive,
    author = "Moulson, Matthew",
    title = "{Experimental determination of $V_{us}$ from kaon decays}",
    eprint = "1704.04104",
    archivePrefix = "arXiv",
    primaryClass = "hep-ex",
    doi = "10.22323/1.291.0033",
    journal = "PoS",
    volume = "CKM2016",
    pages = "033",
    year = "2017"
}

@article{Cirigliano:2001mk,
    author = "Cirigliano, V. and Knecht, M. and Neufeld, H. and Rupertsberger, H. and Talavera, P.",
    title = "{Radiative corrections to K(l3) decays}",
    eprint = "hep-ph/0110153",
    archivePrefix = "arXiv",
    reportNumber = "UWTHPH-2001-44, IFIC-01-55, CPT-2001-P-4248",
    doi = "10.1007/s100520100825",
    journal = "Eur. Phys. J. C",
    volume = "23",
    pages = "121--133",
    year = "2002"
}

@article{Gao:2022xqz,
    author = "Gao, Rui and Guo, Zhi-Hui and Oller, J. A. and Zhou, Hai-Qing",
    title = "{Axion-meson mixing in light of recent lattice {\ensuremath{\eta}}{\textendash}{\ensuremath{\eta}}' simulations and their two-photon couplings within U(3) chiral theory}",
    eprint = "2211.02867",
    archivePrefix = "arXiv",
    primaryClass = "hep-ph",
    doi = "10.1007/JHEP04(2023)022",
    journal = "JHEP",
    volume = "04",
    pages = "022",
    year = "2023"
}

@article{Bernard:2009zm,
    author = "Bernard, Veronique and Oertel, Micaela and Passemar, Emilie and Stern, Jan",
    title = "{Dispersive representation and shape of the K(l3) form factors: Robustness}",
    eprint = "0903.1654",
    archivePrefix = "arXiv",
    primaryClass = "hep-ph",
    doi = "10.1103/PhysRevD.80.034034",
    journal = "Phys. Rev. D",
    volume = "80",
    pages = "034034",
    year = "2009"
}

@article{Marciano:2004uf,
    author = "Marciano, William J.",
    title = "{Precise determination of |V(us)| from lattice calculations of pseudoscalar decay constants}",
    eprint = "hep-ph/0402299",
    archivePrefix = "arXiv",
    doi = "10.1103/PhysRevLett.93.231803",
    journal = "Phys. Rev. Lett.",
    volume = "93",
    pages = "231803",
    year = "2004"
}

@article{Marciano:1988vm,
    author = "Marciano, W. J. and Sirlin, A.",
    title = "{Electroweak Radiative Corrections to tau Decay}",
    doi = "10.1103/PhysRevLett.61.1815",
    journal = "Phys. Rev. Lett.",
    volume = "61",
    pages = "1815--1818",
    year = "1988"
}

@article{Braaten:1990ef,
    author = "Braaten, Eric and Li, Chong-Sheng",
    title = "{Electroweak radiative corrections to the semihadronic decay rate of the tau lepton}",
    reportNumber = "NUHEP-TH-90-30",
    doi = "10.1103/PhysRevD.42.3888",
    journal = "Phys. Rev. D",
    volume = "42",
    pages = "3888--3891",
    year = "1990"
}

@article{Erler:2002mv,
    author = "Erler, Jens",
    title = "{Electroweak radiative corrections to semileptonic tau decays}",
    eprint = "hep-ph/0211345",
    archivePrefix = "arXiv",
    journal = "Rev. Mex. Fis.",
    volume = "50",
    pages = "200--202",
    year = "2004"
}

@article{Antonelli:2013usa,
    author = "Antonelli, Mario and Cirigliano, Vincenzo and Lusiani, Alberto and Passemar, Emilie",
    title = "{Predicting the $\tau$ strange branching ratios and implications for $V_{us}$}",
    eprint = "1304.8134",
    archivePrefix = "arXiv",
    primaryClass = "hep-ph",
    reportNumber = "LA-UR-13-22949",
    doi = "10.1007/JHEP10(2013)070",
    journal = "JHEP",
    volume = "10",
    pages = "070",
    year = "2013"
}

@article{Flores-Baez:2013eba,
    author = "Flores-Ba{\'e}z, F. V. and Morones-Ibarra, J. R.",
    title = "{Model Independent Electromagnetic corrections in hadronic $\tau$ decays}",
    eprint = "1307.1912",
    archivePrefix = "arXiv",
    primaryClass = "hep-ph",
    doi = "10.1103/PhysRevD.88.073009",
    journal = "Phys. Rev. D",
    volume = "88",
    number = "7",
    pages = "073009",
    year = "2013"
}

@article{emcee,
   author = {{Foreman-Mackey}, D. and {Hogg}, D.~W. and {Lang}, D. and {Goodman}, J.},
    title = {emcee: The MCMC Hammer},
  journal = {PASP},
     year = 2013,
   volume = 125,
    pages = {306-312},
   eprint = {1202.3665},
      doi = {10.1086/670067}
}

@article{Lewis:2019xzd,
   author = "Lewis, Antony",
   title = "{GetDist: a Python package for analysing Monte Carlo samples}",
   eprint = "1910.13970",
   archivePrefix = "arXiv",
   primaryClass = "astro-ph.IM",
   doi = "10.1088/1475-7516/2025/08/025",
   journal = "JCAP",
   volume = "08",
   pages = "025",
   year = "2025"
}

@Article{         harris2020array,
 title         = {Array programming with {NumPy}},
 author        = {Charles R. Harris and K. Jarrod Millman and St{\'{e}}fan J.
                 van der Walt and Ralf Gommers and Pauli Virtanen and David
                 Cournapeau and Eric Wieser and Julian Taylor and Sebastian
                 Berg and Nathaniel J. Smith and Robert Kern and Matti Picus
                 and Stephan Hoyer and Marten H. van Kerkwijk and Matthew
                 Brett and Allan Haldane and Jaime Fern{\'{a}}ndez del
                 R{\'{i}}o and Mark Wiebe and Pearu Peterson and Pierre
                 G{\'{e}}rard-Marchant and Kevin Sheppard and Tyler Reddy and
                 Warren Weckesser and Hameer Abbasi and Christoph Gohlke and
                 Travis E. Oliphant},
 year          = {2020},
 month         = sep,
 journal       = {Nature},
 volume        = {585},
 number        = {7825},
 pages         = {357--362},
 doi           = {10.1038/s41586-020-2649-2},
 publisher     = {Springer Science and Business Media {LLC}},
 url           = {https://doi.org/10.1038/s41586-020-2649-2}
}

@ARTICLE{2020scipy-nmeth,
  author  = {Virtanen, Pauli and Gommers, Ralf and Oliphant, Travis E. and
            Haberland, Matt and Reddy, Tyler and Cournapeau, David and
            Burovski, Evgeni and Peterson, Pearu and Weckesser, Warren and
            Bright, Jonathan and {van der Walt}, St{\'e}fan J. and
            Brett, Matthew and Wilson, Joshua and Millman, K. Jarrod and
            Mayorov, Nikolay and Nelson, Andrew R. J. and Jones, Eric and
            Kern, Robert and Larson, Eric and Carey, C J and
            Polat, {\.I}lhan and Feng, Yu and Moore, Eric W. and
            {VanderPlas}, Jake and Laxalde, Denis and Perktold, Josef and
            Cimrman, Robert and Henriksen, Ian and Quintero, E. A. and
            Harris, Charles R. and Archibald, Anne M. and
            Ribeiro, Ant{\^o}nio H. and Pedregosa, Fabian and
            {van Mulbregt}, Paul and {SciPy 1.0 Contributors}},
  title   = {{{SciPy} 1.0: Fundamental Algorithms for Scientific
            Computing in Python}},
  journal = {Nature Methods},
  year    = {2020},
  volume  = {17},
  pages   = {261--272},
  adsurl  = {https://rdcu.be/b08Wh},
  doi     = {10.1038/s41592-019-0686-2},
}

@Article{Hunter:2007,
  Author    = {Hunter, J. D.},
  Title     = {Matplotlib: A 2D graphics environment},
  Journal   = {Computing in Science \& Engineering},
  Volume    = {9},
  Number    = {3},
  Pages     = {90--95},
  publisher = {IEEE COMPUTER SOC},
  doi       = {10.1109/MCSE.2007.55},
  year      = 2007
}

@article{Zielinski:2025wfa,
    author = "Zieli{\'n}ski, Marcin and Gatto, Corrado",
    title = "{The REDTOP Experiment: an \(\eta /\eta ^{\prime }\) Factory to Explore Dark Matter and Physics Beyond the Standard Model}",
    eprint = "2509.26552",
    archivePrefix = "arXiv",
    primaryClass = "hep-ex",
    doi = "10.5506/APhysPolBSupp.18.4-A5",
    journal = "Acta Phys. Polon. Supp.",
    volume = "18",
    number = "4",
    pages = "4-A5",
    year = "2025"
}

\end{document}